\documentclass[sort&compress,5p,times]{elsarticle}
\pdfoutput=1
\journal{Solid State Ionics}

\usepackage[figuresright]{rotfloat}
\usepackage[english]{babel}
\usepackage[T1]{fontenc}
\usepackage[utf8]{inputenc} 
\usepackage{amsmath}%
\usepackage{mathtools}
\usepackage{amsfonts}%
\usepackage{amssymb}%
\usepackage[ampersand]{easylist}
\usepackage{graphicx}%
\usepackage{float}
\usepackage{multirow}
\usepackage{multicol}
\usepackage{subcaption} 
\usepackage{setspace}
\usepackage{upgreek}
\usepackage{units} 		
\usepackage{lipsum} 	
\usepackage{wrapfig}
\usepackage{afterpage}
\usepackage{epsfig}
\usepackage{placeins}
\usepackage{tablefootnote}	
\usepackage[flushleft]{threeparttable}

\newcommand{\tr}[1]{\text{Tr}\left({#1}\right)}

\newcommand{\mat}[1]{\mathsf{#1}}
\newcommand{\eq}{\begin{equation}}			
\newcommand{\eeq}{\end{equation}}
\newcommand{\eqs}{\begin{equation*}}		
\newcommand{\eeqs}{\end{equation*}}
\newcommand{\ealign}{\begin{align}}
\newcommand{\eealign}{\end{align}}
\newcommand{\ealigns}{\begin{align*}}		
\newcommand{\eealigns}{\end{align*}}

\renewcommand{\vec}{\mathbf}
\newcommand{\etal}{\textit{et al.~}} 

\newcommand{\eps}{\varepsilon}

\renewcommand{\rm}{\mathrm}
\newcommand{\matris}[1]{$\mathsf{#1}$}

\newcommand{\rd}{\ensuremath{\mathrm{d}}}

\newcommand{\PD}[2]{\frac{\partial#1}{\partial#2}}

\usepackage[usenames,dvipsnames,svgnames,table]{xcolor}

{}
{}
{}

\newcommand{\bzo}{BaZrO$_3$}

\begin{document}
\selectlanguage{english}

\begin{frontmatter}

\title{Size and shape of oxygen vacancies and protons in acceptor-doped barium zirconate}
\author{Erik Jedvik}
\author{Anders Lindman}
\author{Magnús \TH ór Benediktsson}
\author{Göran Wahnström\corref{cor1}}
\cortext[cor1]{Corresponding author E-mail address: goran.wahnstrom@chalmers.se}
\address{Department of Applied Physics, Chalmers University of Technology, SE-412 96 Göteborg, Sweden}

\begin{abstract}
The defect induced chemical expansion in acceptor-doped barium zirconate is investigated using density-functional theory (DFT) calculations. The two defect species involved in the hydration reaction, the $+2$ charged oxygen vacancy and the proton interstitial forming a hydroxide ion, are considered both as free defects and in association with the dopants Y, In, Sc and Ga. The defect induced strain tensor $\lambda$ is introduced, which provides a natural generalisation of the ordinary chemical expansion to three dimensions and to anisotropic distortions. Both the addition of a vacancy and a proton cause anisotropic distortions and a net contraction of the lattice, indicating that both the vacancy and the hydroxide ion are smaller than the oxygen ion. The contraction is considerably larger for the vacancy and the net effect in hydration, when a vacancy is filled and two protons are added, is an expansion, consistent with the experimental findings. The effect of the dopants on the chemical expansion in 
hydration is found to be quite small, even if it is assumed that both the vacancy and the proton are fully associated with a dopant atom in the lattice.
\end{abstract}


\end{frontmatter}

\section{Introduction\label{sec:intro}}

Over the past three decades acceptor-doped perovskites have been studied extensively for the use as proton conducting electrolytes in intermediate temperature fuel cells, electrolyzers, etc.~\cite{norby_proton_2009,Kreuer_AnnRevMatRes2003}. One of the most promising materials is BaZrO$_3$, which combines high bulk proton conductivity together with chemical stability towards CO and CO$_2$ \cite{Fabbri_AdvMat_2012}.

Protons are incorporated into the perovskite structure through the hydration of oxygen vacancies, which are formed due to the acceptor doping in order to maintain charge neutrality. In Kröger-Vink notation the hydration reads as 
\eq
\label{eq:hydration}
\text{H}_2\text{O}(\text{g}) + \text{v}_{\text{O}}^{\bullet\bullet} + \text{O}_{\text{O}}^{\text{x}} \rightleftharpoons 2\text{OH}_{\text{O}}^{\bullet}
\eeq

It has been seen experimentally that the hydration of acceptor-doped BaZrO$_3$ is associated with a volume expansion \cite{Kreuer_AnnRevMatRes2003,EricssonSSI225_2012,HiraiwaJAmCeramSoc96_2013,han_dopant_2014,KneeGrande_JACS_97} that puts the material under mechanical stress and can lead to micro-cracking of the dense electrolyte and ultimately deterioration of the cell performance. Chemical expansion due to oxidation has been reported for other well-known ionic conductors, such as CeO$_{2-\delta}$ \cite{MarrocchelliAFM12,marrocchelli_charge_2012} and La$_{1-x}$Sr$_x$Co$_{1-y}$Fe$_{y}$O$_{3-\delta}$ \cite{Adler_JAmCeramSoc84_2001,Bishop_JACS_93}. In these materials the expansion is related to two phenomena: (1) the formation of oxygen vacancies and (2) the reduction of cations, which change the ionic radius of those ions, e.g.~$\text{Ce}^{4+}\rightarrow\text{Ce}^{3+}$. The latter effect is not present in BaZrO$_3$ as the barium and zirconium cations are isovalent. Thus, the hydration induced volume expansion 
should be more directly related to the size of oxygen vacancies and proton interstitials (hydroxide ions).
  
The size of the hydroxide ion is well known in terms of the Shannon ionic radius, and it is slightly smaller than the oxygen ion \cite{ShannonActaCrystA32_1976}. However, the size of the charged oxygen vacancy is not as well defined. Suggestions have been made that the formation of a vacancy should lead to an expansion due to the electrostatic repulsion between the cations located closest to the vacancy \cite{Atkinson_SolidStateIonics_2000,Bishop_JACS_93}. On the other hand, studies of CeO$_2$ and ZrO$_2$ indicate that the vacancy is smaller than the oxygen ion \cite{hong_lattice_1995,MarrocchelliAFM12,marrocchelli_charge_2012,Chatzichristodoulou_size_2014}. 

In this paper we study the size of the +2 charged oxygen vacancy and the proton interstitial in acceptor-doped BaZrO$_3$ using density-functional theory (DFT). The size of the two defects is determined in terms of the defect induced strain tensor. This quantity does not only predict the lattice expansion, \emph{''the size''} of the defect, but also anisotropic expansions, \emph{''the shape''} of the defect. Furthermore, the implications of the results on the chemical expansion due to hydration are discussed both in the dilute limit of freely moving defects and as trapped defects in association with different dopants.

\section{Theoretical formalism\label{sec:formalism}}

\subsection{Chemical expansion coefficient}

To quantify the chemical expansion induced in a crystal upon formation of point defects we start from a classical thermodynamics point of view \cite{Adler_JAmCeramSoc84_2001}. We consider the volume, $V$, to depend on temperature, $T$, pressure, $P$, and defect concentration $x_k$ of type $k$
\eqs
V = V(T,P,\lbrace x_k \rbrace)
\eeqs
The unit for the defect concentration, $x_k$, is the number of defects per primitive unit cell (one formula unit \bzo) and the volume of one primitive unit cell is denoted $\Omega_c$. In analogy with the thermal expansion coefficient $\alpha = V^{-1}(\partial V/ \partial T)_{P,x_k}$ and compressibility $\kappa = -V^{-1}(\partial V/ \partial P)_{T,x_k}$ we define the \emph{chemical expansion coefficient} as \cite{Adler_JAmCeramSoc84_2001}
\eq
\beta_k = \frac{1}{V} \left( \PD{V}{x_k} \right)_{P,T,x_{k' \neq k}}
\eeq
The coefficient $\beta_k$ corresponds to the chemical \emph{volume} expansion of the material and in general we have $\beta_k = \beta_k (T,P,\lbrace x_k \rbrace)$. 
The differential for the volume can be written as
\eq
\frac{\rd V}{V} = \alpha \rd T - \kappa \rd P +\sum\limits_k \beta_k \rd x_k
\label{eq:V_differential}
\eeq

To describe the expansion in a solid material including aniso\-tropy, we generalise the concept of a scalar volume and introduce the elastic strain tensor $\epsilon$.
The volume expansion, or volume dilatation, of a material under assumption of small strain is given by the trace of the strain tensor $\epsilon$ \cite{CallenThermo,Kittel}
\eqs
\frac{\rd V}{V}  = \tr{\epsilon}
\eeqs
and the chemical expansion coefficient can thus be written as
\eq
\beta_k =  \tr{\PD{\epsilon}{x_k}}_{P,T,x_{k' \neq k}}
\label{eq:beta_definition}
\eeq

\subsection{Strain Tensor}

In order to derive an expression for the strain tensor we consider the deformation of a homogeneous anisotropic elastic medium. Let
\eqs
\vec{a}_i = \mat{L_0}\vec{e}_i
\eeqs
be the basis vectors defining the undeformed crystal and \matris{L_0} the one-to-one matrix transformation that maps the cartesian coordinates $\vec{e}_i$ onto the crystal $\vec{a}_i$.
After the deformation the crystal is defined by the new basis vectors
\eqs
\vec{a'}_i = \mat{L}\vec{e}_i = \mat{L}\mat{L_0^{-1}} \vec{a}_i
\eeqs
The location of a point, $P = (p_1, p_2, p_3)$, in the undeformed crystal can then be written as
\eqs
\vec{r} = \sum_i p_i \, \vec{a_i}
\eeqs 
and after the deformation
\eqs
\vec{r'} = \sum_i p_i \, \vec{a_i}'
\eeqs 
The displacement of the point is thus
\eqs
\vec{u} = \vec{r'} - \vec{r} = \sum_i p_i \, (\vec{a}_i' - \vec{a}_i) = \left( \mat{L}\mat{L_0^{-1}} - \mat{I} \right) \vec{r}
\eeqs 
where the $\mat{I}$ is the identity matrix and the quantity $\left( \mat{L} \mat{L_0^{-1}} - \mat{I} \right)$ is recognised as the displacement gradient in the Lagrangian description of continuum mechanics \cite{Holzapfel}. In tensor notation the displacement gradient can be written as
\eqs
\left( \mat{L}\mat{L_0^{-1}} - \mat{I} \right)_{ij} = \PD{u_i}{r_j}
\eeqs 
and describes the deformation of the crystal \cite{CallenThermo}.
In addition to strain, the displacement gradient can also include rotations. Rotations should be eliminated, since they do not alter the internal relationships between the atoms and are of no thermodynamic significance. In general this can be done by means of a polar decomposition \cite{Holzapfel}, but for small rotations it is sufficient to add the transpose and take the mean. 
Hence the strain tensor is given by
\eq 
\epsilon_{ij} = \tfrac{1}{2} \left( \PD{u_i}{r_j} + \PD{u_j}{r_i}\right)
\eeq 
or in matrix notation
\eq 
\epsilon = \tfrac{1}{2} \left( \mat{L}\mat{L_0^{-1}} + \left(\mat{L_0^{-1}} \right)^T \mat{L}^T \right) - \mat{I} 
\eeq 
If no rotations are present the strain tensor $\epsilon$ equals the displacement gradient
\eq
\epsilon = \left( \mat{L} - \mat{L_0} \right)\mat{L_0^{-1}}
\label{eq:epsilon_definition}
\eeq
This expression is the natural generalisation of the one dimensional strain $ \eps = \Delta a / a_0$ to higher dimensions and is similar to the one proposed by Centoni \emph{et al.} \cite{Centoni_PRB_2005}.

\subsection{Defect induced strain tensor\label{sec:defect_induced_strain_tensor}}

Consider now the effect of a single defect introduced in a volume $V_0$.
The \emph{defect induced strain tensor} $\lambda$ is defined as
\eq
\lambda = \frac{1}{x_d} \epsilon
\label{eq:lambda_definition}
\eeq
where $x_d = \Omega_c/V_0$ is the defect concentration with $\Omega_c$ the volume of the primitive unit cell.
Hence the chemical expansion coefficient in Eq. \eqref{eq:beta_definition} can be written as 
\eq
\beta = \tr{\lambda}
\label{eq:beta_definition_2}
\eeq
The tensor $\lambda$ provides a natural generalisation of the ordinary chemical expansion coefficient to anisotropic deformations. It is identical to the $\lambda$-tensor introduced by Nowick and Berry \cite{NowicBerry}, and directly related to the defect strain tensor $\Lambda = \Omega_c \lambda$ used by Freedman \emph{et al.} \cite{Freedman_PRB_2009}.

\section{Computational method\label{sec:method}}

Electronic structure calculations were performed using density-functional theory (DFT) as implemented in the VASP software (version 5.3.3) \cite{VASP1,VASP2,VASP3,VASP4}, which uses a plane-wave basis set with periodic boundary conditions in all directions. 
The generalised gradient approximation (GGA) PBE \cite{PBE} was used for the exchange-correlation functional and the projector augmented wave method (PAW) \cite{VASP5,VASP6} to describe the ion-electron interaction.
Two supercell sizes were used, consisting of $4\times4\times4$ and $2\times2\times2$ primitive unit cells,
which implies defect concentrations of 1/64 and 1/8 respectively. 
The corresponding $\vec{k}$-point sampling was $2\times2\times2$ and $4\times4\times4$ using the Monkhorst-Pack scheme.
The energy cutoff was set to 520 eV.
To compensate for the electrical charge of the defects a homogeneous background charge was added.

Calculating the strain requires high accuracy. This is due to that the strain, which is a small quantity, is obtained as the difference between two similar values (see Eq. \eqref{eq:epsilon_definition}).
The electronic convergence criterion was set to $10^{-7}$ eV and the cell was allowed to relax in both size and shape until the forces were less than $10^{-3}$ eV/Å. 

The calculations were performed according to the following procedure. An ideal bulk \bzo\ supercell was constructed and relaxed with resulting supercell parameters \matris{L_0}. A single defect was then introduced into the supercell followed by atomic structure relaxation where the supercell size and shape were allowed to change, giving the defective supercell parameters defined by $\mat{L}$. The strain tensor $\epsilon$ was then obtained from $\mat{L}$ and $\mat{L}_0$ using Eq. \eqref{eq:epsilon_definition}. From here the defect induced strain tensor $\lambda$ and the chemical expansion coefficient $\beta$ were determined using Eqs. \eqref{eq:lambda_definition} and \eqref{eq:beta_definition_2}. 
For the doped systems we first relaxed the structure with the dopant alone, which gave the corresponding chemical expansion. 
A defect was then added followed by further relaxation. In this case the doped system served as \matris{L_0}.

\section{Results\label{sec:results}}

In this work we consider the two defects associated with the hydration reaction in Eq. \eqref{eq:hydration}: the +2 charged oxygen vacancy ($\text{v}_{\text{O}}^{\bullet\bullet}$) and the proton ($\text{OH}_{\text{O}}^{\bullet}$) \cite{sundell_thermodynamics_2006,bjorketun_structure_2007}.
The computational approach makes it possible to study these defects without having to consider the dopant atom explicitly by instead adding an appropriate homogeneous background charge \cite{freysoldt_first-principles_2014}.
However, we have also studied the effect of vacancy-dopant and proton-dopant association by explicitly considering four different dopants: Y, In, Sc, and Ga.

The ground state for the defect free system is found to be cubic (Pm$\bar{3}$m) with lattice constant $a_0 = \unit[4.236]{\text{Å}}$. This is somewhat larger than the experimental value of \unit[4.19]{Å} \cite{levin_phase_2003} but is in line with previous GGA results in the literature \cite{Gomez_JChemPhys_123_2005,shi_first-principles_2005,bjorketun_kinetic_2005,bjorheim_combined_2010}.

\subsection{Oxygen vacancy}
\begin{figure}[!tb]
\begin{center}
\begin{subfigure}[t]{0.06\textwidth}
\centering
\includegraphics[width=\textwidth]{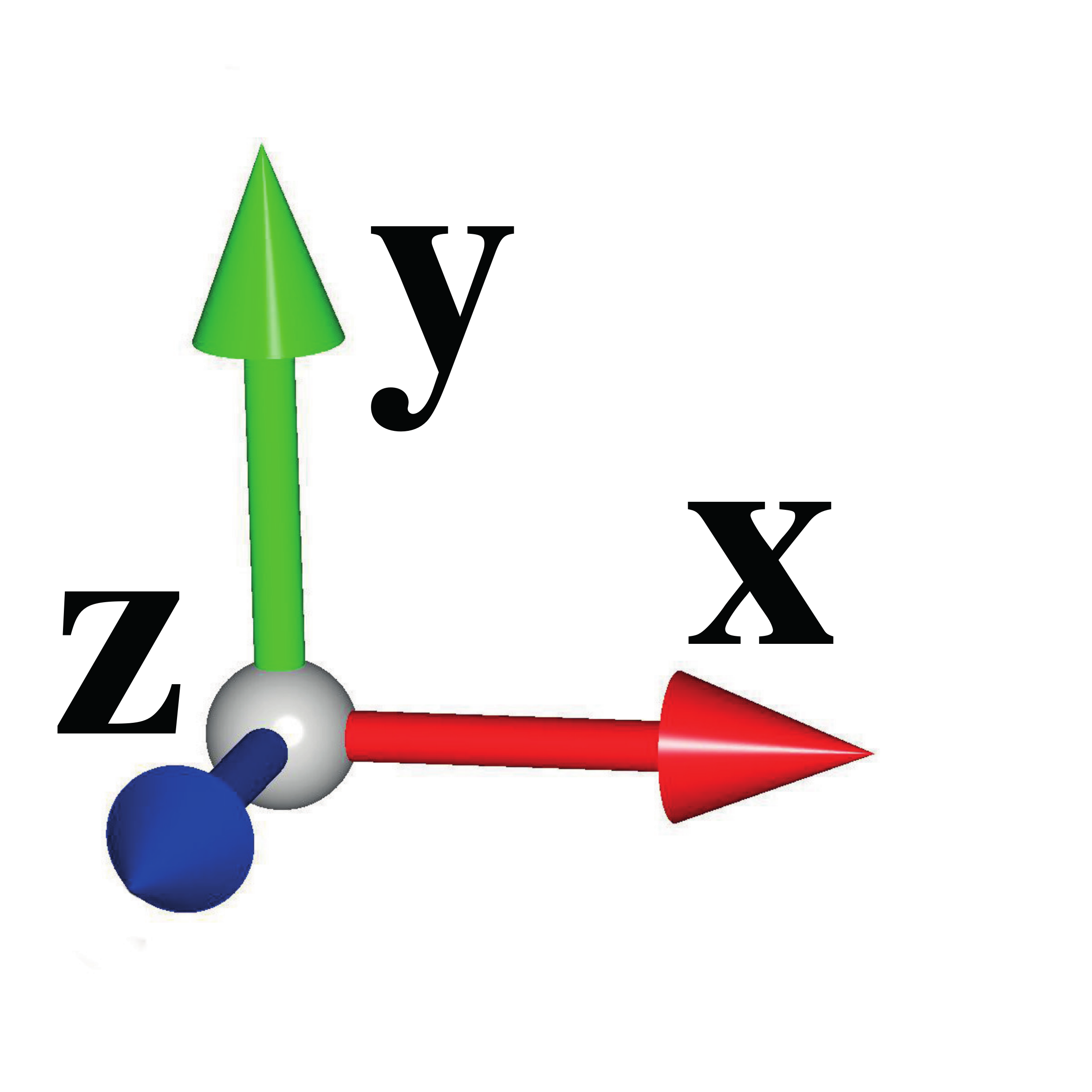}
\end{subfigure}
\begin{subfigure}[t]{0.18\textwidth}
\centering
\includegraphics[width=\textwidth]{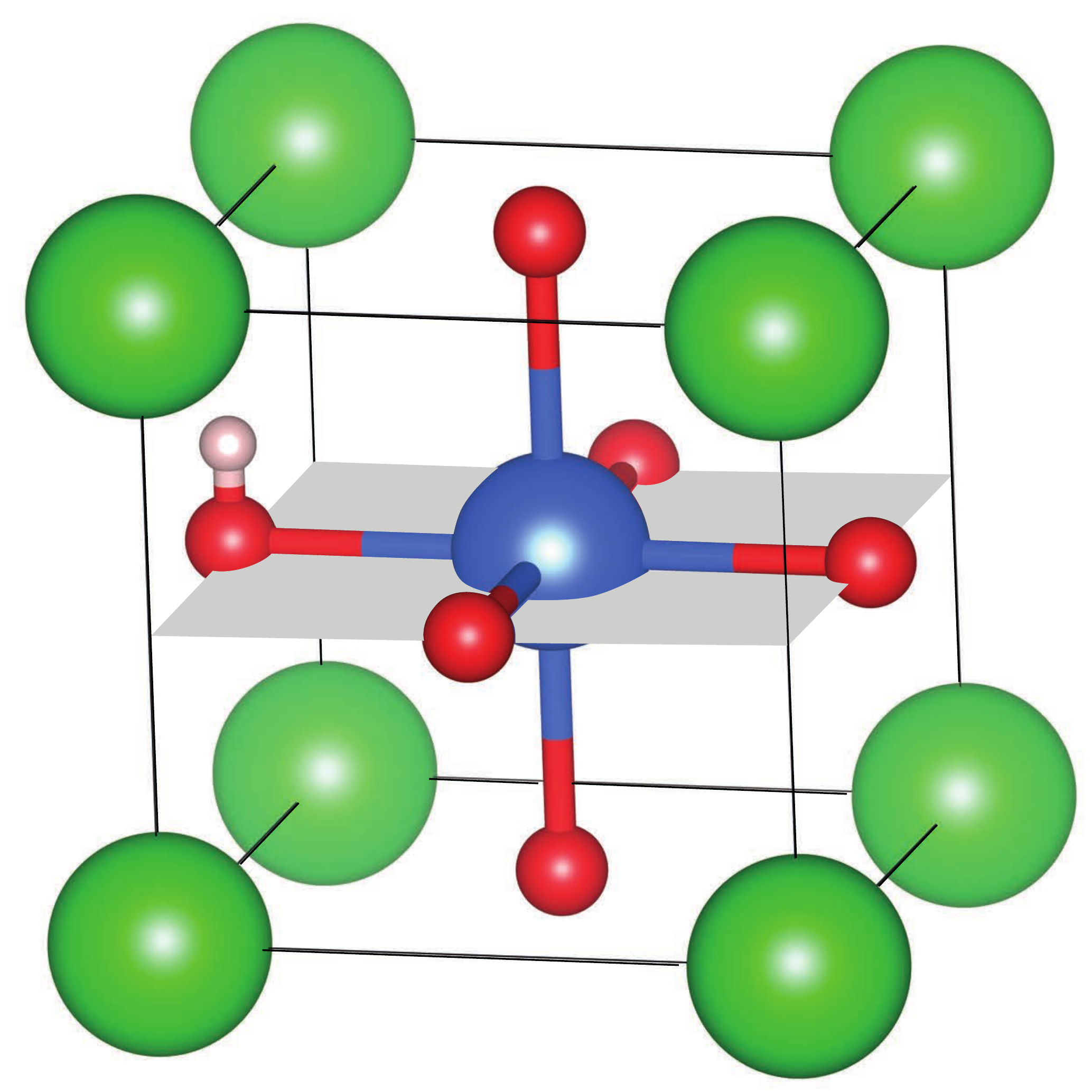}
\caption{ }
\label{fig:BZO_cell_plane2}
\end{subfigure}
\begin{subfigure}[t]{0.18\textwidth}
\centering
\includegraphics[width=\textwidth]{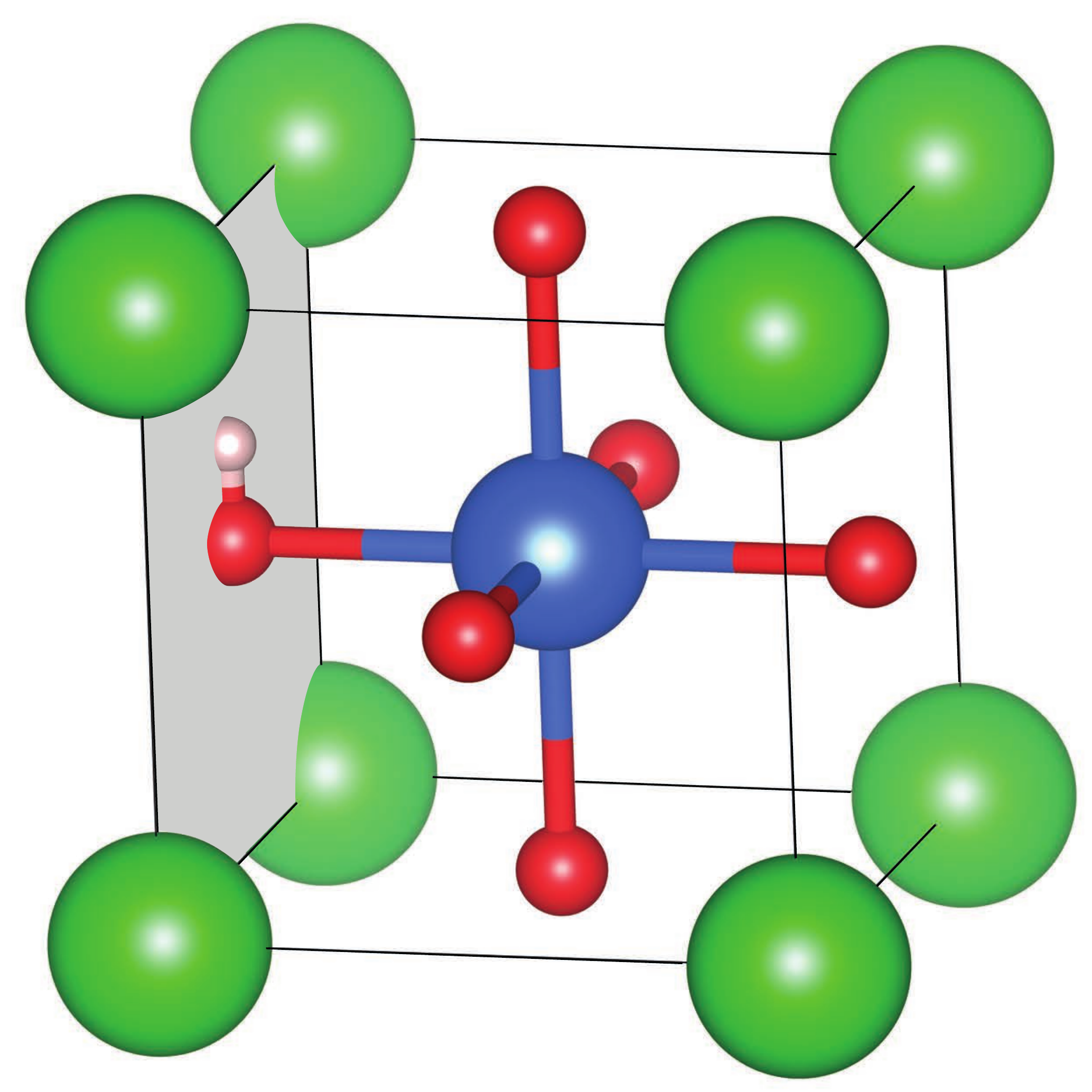}
\caption{  }
\label{fig:BZO_cell_plane1}
\end{subfigure}
\caption{Illustrations of (a) the ZrO$_2$-plane [010] and (b) the BaO-plane [100]. The red, blue, green and white atoms correspond to oxygen, zirconium, barium and hydrogen, respectively.  The added proton shows the defect lattice site as well as the orientation of the hydroxide ion.}
\label{fig:BZO_cell_planes}
\end{center}
\end{figure}

\begin{figure}[!tb]
\begin{center}
\begin{subfigure}[t]{0.4\textwidth}
\centering
\includegraphics[width=\textwidth]{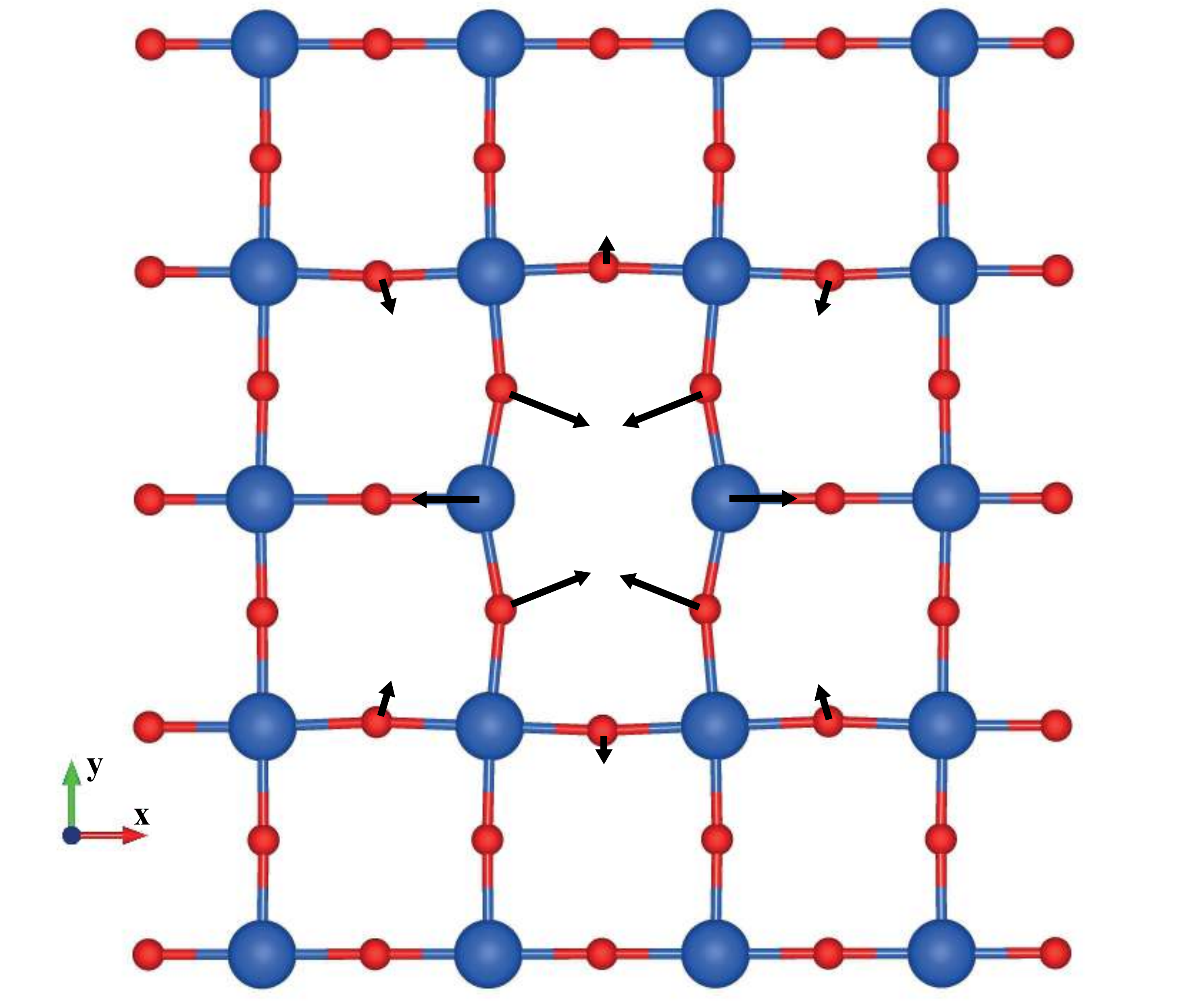}
\caption{  }
\label{fig:latticeDistortionsZrV}
\end{subfigure}
\begin{subfigure}[t]{0.4\textwidth}
\centering
\includegraphics[width=\textwidth]{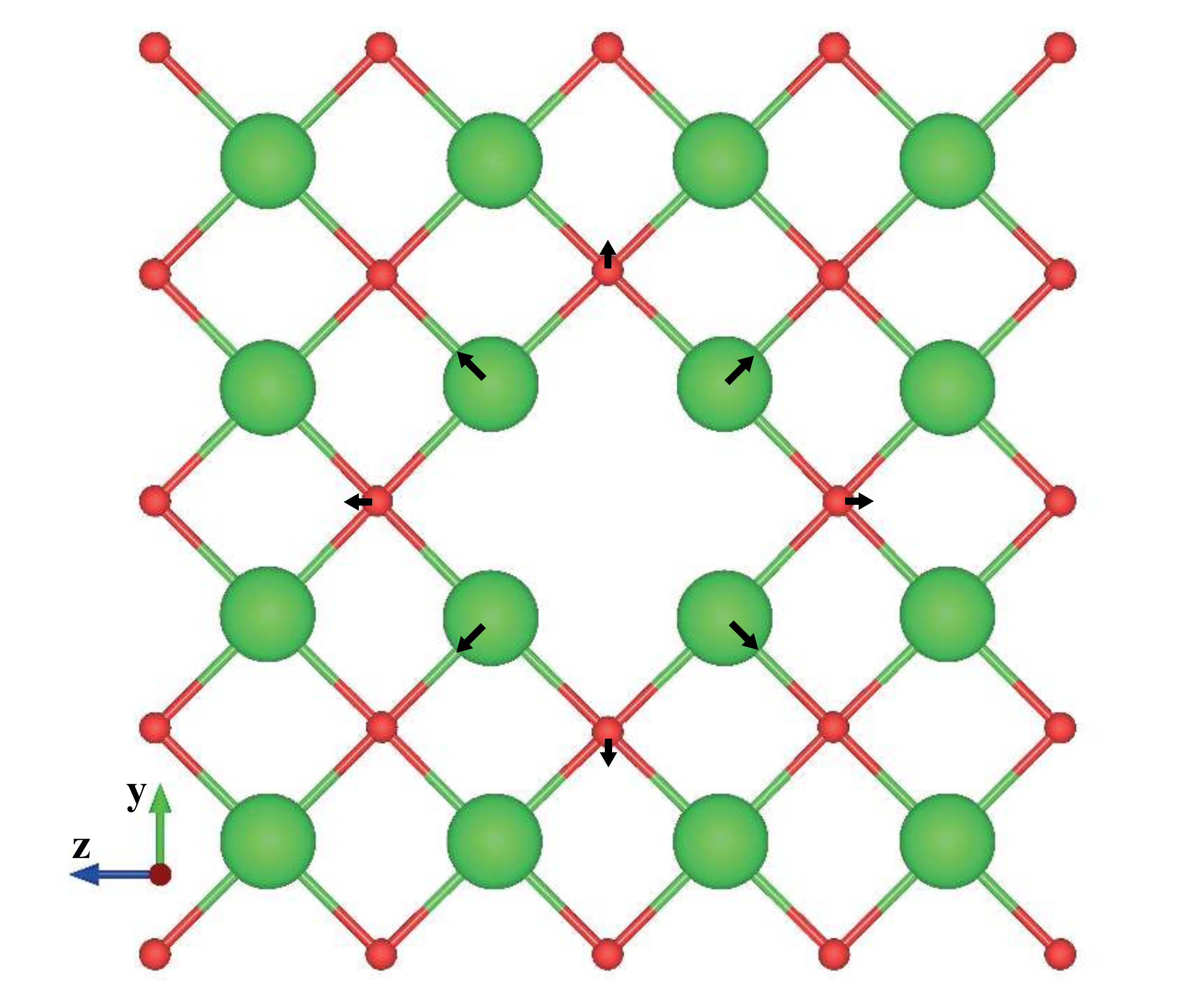}
\caption{  }
\label{fig:latticeDistortionsBaV}
\end{subfigure}
\caption{Visualisation of the lattice distortions for an oxygen vacancy in (a) the [010] ([001]) plane (ZrO$_2$) and (b) the [100] plane (BaO). Colours follow the same convention as in Fig. \ref{fig:BZO_cell_planes} and the arrows indicate the displacements larger than \unit[0.05]{Å} in Tab. \ref{tab:distortions}.}
\label{fig:latticeDistortionsV}
\end{center}
\end{figure}

First we consider the oxygen vacancy. Fig. \ref{fig:BZO_cell_planes} illustrates the ZrO$_2$ and BaO lattice planes, [010] and [100] respectively, which intersect at the defect. The [001] ZrO$_2$-plane is not explicitly shown. The ionic displacements in these planes due to the formation of an oxygen vacancy are shown in Fig. \ref{fig:latticeDistortionsV}. Fig. \ref{fig:latticeDistortionsZrV} corresponds to distortions in the [010] and [001] planes (ZrO$_2$) which are equivalent due to symmetry, while Fig. \ref{fig:latticeDistortionsBaV} corresponds the [100] plane (BaO).

The ionic displacements are listed in Tab. \ref{tab:distortions} and are determined with respect to the defect lattice site, the lattice site where the defect is created. In general, the displacements can be understood from electrostatic considerations. The vacancy, having an effective charge of $+2$, repels the metal cations and attracts the negatively charged oxygen ions. For the nearest neighbouring zirconium and barium ions the displacements are \unit[0.17]{Å} and \unit[0.10]{Å} respectively, and directed away from the vacancy, while the nearest neighbouring oxygen ions are displaced \unit[0.22]{Å} towards the vacancy. Ions further away from the vacancy are displaced as well, although to a lesser extent.  

\begin{table}[!tb]
\caption{Displacements of ions relative to the defect lattice site. The first three columns state the number of the coordination shell (CS) in the ideal cell, the distance to the defect lattice site ($d$) and the atomic species, respectively. The fourth and sixth column indicate the coordination number (CN) of each ion and the fifth and seventh the length of the displacement vector ($\Delta d$). The sign indicate whether the radial component of the displacement vector is towards $(-)$ or away from $(+)$ the defect.}
\label{tab:distortions}
\begin{center}
\begin{tabular}{@{\extracolsep{\stretch{1}}}crcclcl}\hline

\multicolumn{3}{c}{ }	&  \multicolumn{2}{c}{Vacancy}	&\multicolumn{2}{c}{Proton}	\\ \cline{4-5} \cline{6-7}
CS	&   $d$ ($a_0$)	&	Atom	&	CN	&	$\Delta d$	 (Å)	&	CN	& 	$\Delta d$ (Å)	\\ \hline	
1	&$1/2$		&	Zr	&	2	&	0.173	$(+)$	&	2	&	0.155	$(+)$	\\
2	&$\sqrt{2}/2$	&	Ba	&	4	&	0.096	$(+)$	&	2	&	0.198	$(+)$	\\
	&		&		&		&			&	2	&	0.059	$(-)$	\\
	&		&	O	&	8	&	0.220	$(-)$	&	4	&	0.063	$(+)$	\\
	&		&		&		&			&	2	&	0.085	$(-)$	\\
	&		&		&		&			&	2	&	0.214	$(-)$	\\
3	&1		&	O	&	2	&	0.000	$(+)$	&	2	&	0.159	$(+)$	\\
	&		&		&	4	&	0.072	$(+)$	&	1	&	0.175	$(+)$	\\
	&		&		&		&			&	1	&	0.107	$(-)$	\\
	&		&		&		&			&	2	&	0.093	$(-)$	\\
4	&$\sqrt{5}/2$	&	Zr	&	8	&	0.018	$(-)$	&	4	&	0.059	$(+)$	\\
	&		&		&		&			&	2	&	0.040	$(+)$	\\
	&		&		&		&			&	2	&	0.060	$(-)$	\\
5	&$\sqrt{6}/2$	&	Ba	&	8	&	0.023	$(-)$	&	4	&	0.065	$(+)$	\\
	&		&		&		&			&	4	&	0.068	$(-)$	\\
	&		&	O	&	8	&	0.084	$(+)$	&	4	&	0.092	$(+)$	\\
	&		&		&		&			&	4	&	0.053	$(+)$	\\
	&		&		&	8	&	0.023	$(-)$	&	4	&	0.040	$(+)$	\\
	&		&		&		&			&	4	&	0.073	$(-)$	\\
6	&$\sqrt{2}$	&	O	&	4	&	0.026	$(-)$	&	2	&	0.027	$(+)$	\\
	&		&		&	8	&	0.094	$(-)$	&	2	&	0.047	$(-)$	\\
	&		&		&		&			&	4	&	0.036	$(+)$	\\
	&		&		&		&			&	2	&	0.037	$(-)$	\\
	&		&		&		&			&	2	&	0.065	$(-)$	\\
\hline
\end{tabular}
\end{center}
\end{table}

The displacement of the outermost oxygen ions in Fig. \ref{fig:latticeDistortionsV} indicates the displacement of the boundary of the simulation cell and are related to the defect induced strain tensor $\lambda$. The oxygen ions are all displaced towards the vacancy leading to a defect induced strain tensor for the oxygen vacancy of 
\eqs
\lambda_{\text{v}} = 
\begin{pmatrix}
-0.038  & 0 		& 0     \\
0	& -0.096	& 0     \\
0	& 0		& -0.096     \\
\end{pmatrix}
\label{eq:strainTensorV}
\eeqs
The components along the diagonal correspond to the usual linear expansion $\eps = \Delta a/a_0$ in the different directions divided by the defect concentration. All components are negative indicating a contraction of the lattice in all directions. The first component is different from the second and third components, which are equal due to symmetry, resulting in an anisotropic strain. The contraction is smallest in the $x$-direction, the same direction in which the nearest neighbouring zirconium ions are being displaced. Since all components of $\lambda_{\text{v}}$ are negative it follows that the chemical expansion coefficient is negative as well. The value we obtain is
\eqs
\beta_{\text{v}}= -0.230
\eeqs
which indicates that there is a decrease in volume upon formation of a vacancy, or in other words, the oxygen vacancy is smaller than an oxygen ion. 

\subsection{Proton}
\begin{figure}[!tb]
\begin{center}
\begin{subfigure}[t]{0.38\textwidth}
\centering
\includegraphics[width=\textwidth]{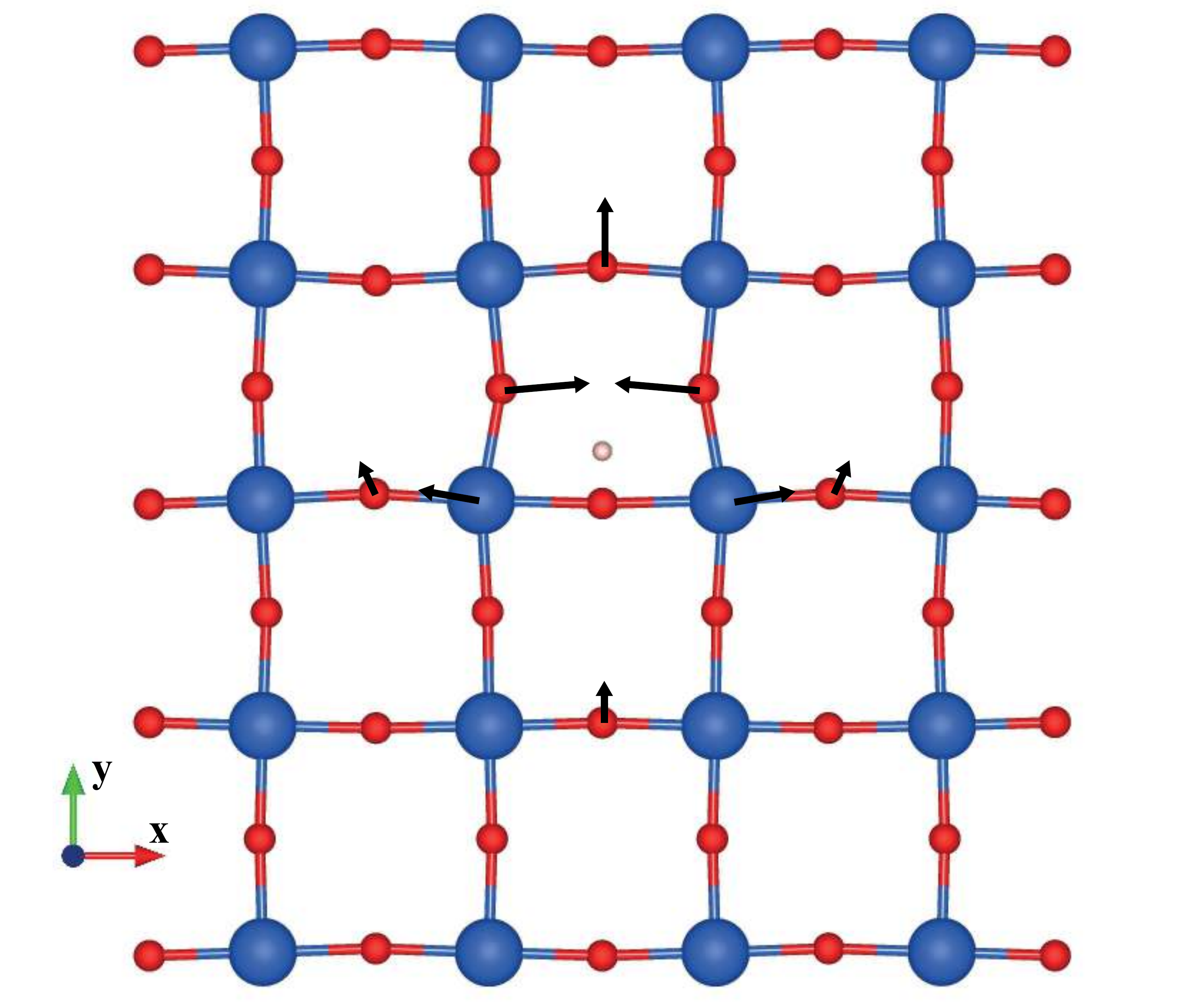}
\caption{  }
\label{fig:latticeDistortionsZrP}
\end{subfigure}
\begin{subfigure}[t]{0.38\textwidth}
\centering
\includegraphics[width=\textwidth]{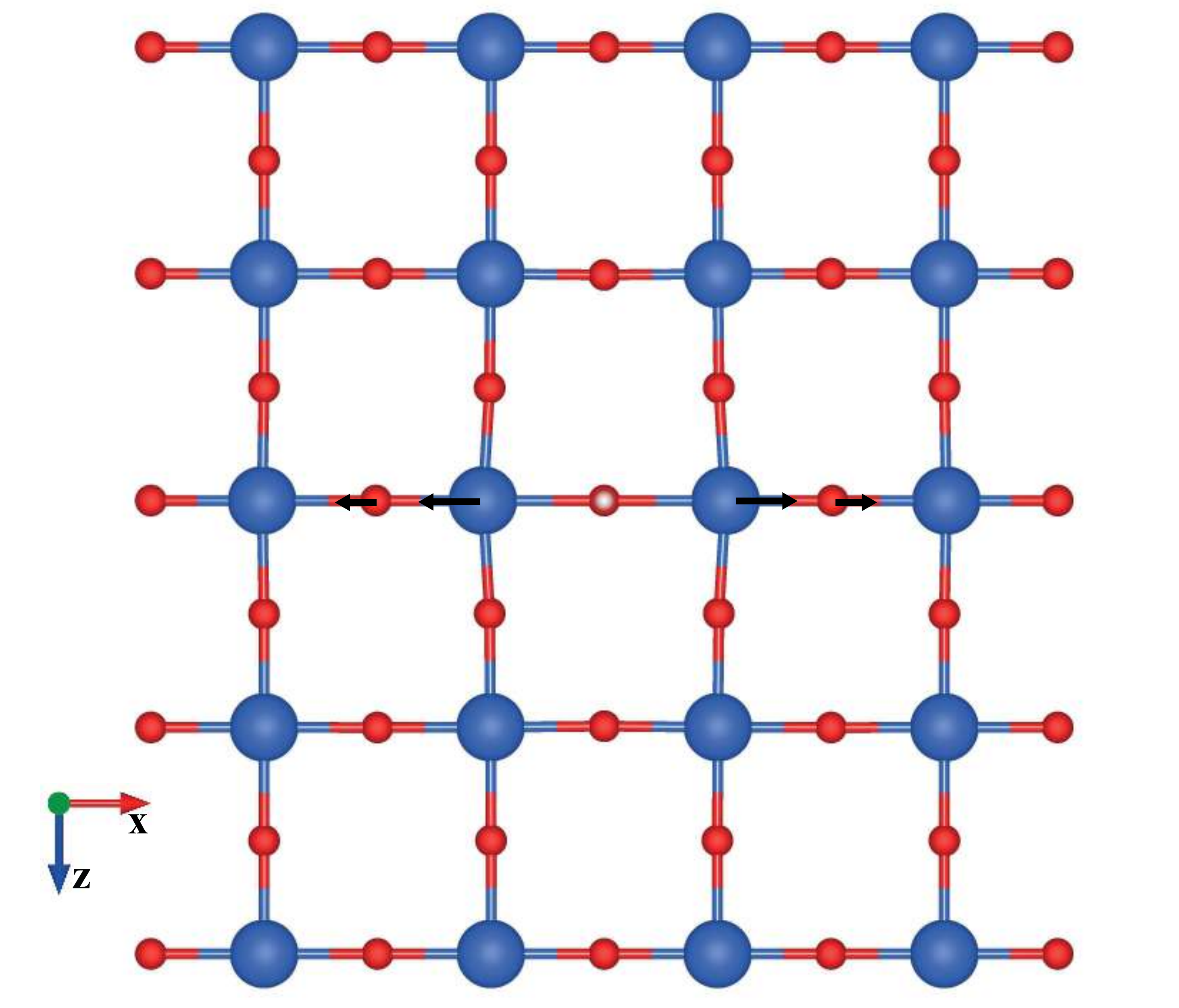}
\caption{  }
\label{fig:latticeDistortionsZrPperp}
\end{subfigure}
\begin{subfigure}[t]{0.38\textwidth}
\centering
\includegraphics[width=\textwidth]{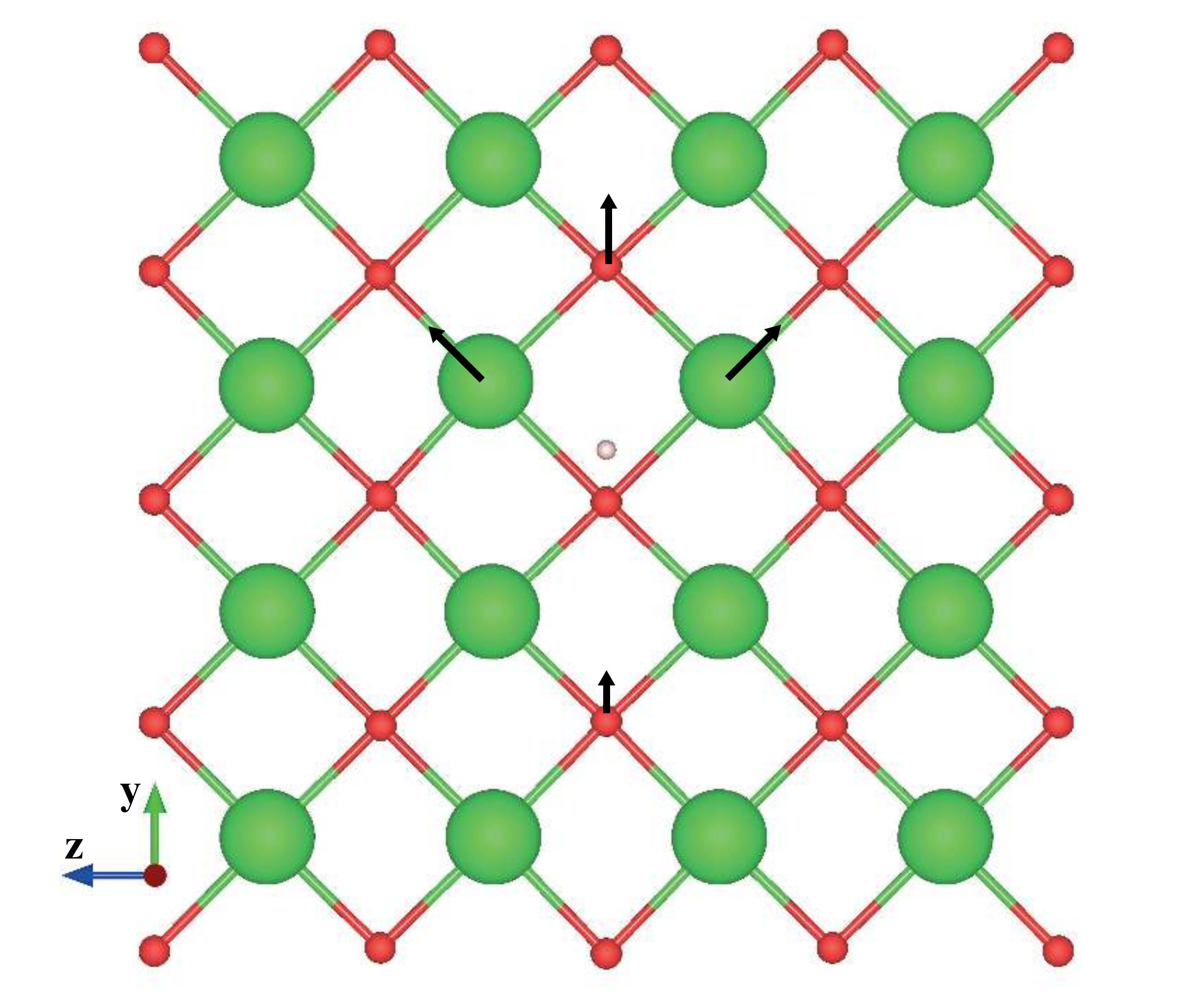}
\caption{  }
\label{fig:latticeDistortionsBaP}
\end{subfigure}
\caption{Visualisation of the lattice distortions for a proton interstitial in (a) the [001] plane (ZrO$_2$), (b) the [010] plane (ZrO$_2$) and (c) the [100] plane (BaO). Colours follow the same convention as in Fig. \ref{fig:BZO_cell_planes} and the arrows indicate the displacements larger than \unit[0.10]{Å}.}
\label{fig:latticeDistortionsP}
\end{center}
\end{figure}

The distortions induced by the proton interstitial are shown in Fig. \ref{fig:latticeDistortionsP}. The proton reduces the symmetry with respect to the vacancy and consequently the distortions in the [010] and [001] planes (ZrO$_2$) are no longer equivalent.  

Tab. \ref{tab:distortions} summarises the proton induced local distortions on individual ions. Also in this case the displacements are determined with respect to the defect lattice site. Similar to the oxygen vacancy the nearest neighbouring zirconium ions are repelled by the proton causing a displacement of \unit[0.16]{Å}. The four closest barium ions are also repelled, although two of them are substantially more displaced than the others (\unit[0.20]{Å} and \unit[0.06]{Å} respectively) due to the asymmetric nature of the hydroxide ion along the $y$-direction. For the nearest neighbouring oxygen ions the picture is somewhat different compared to the oxygen vacancy. The ions in the [001] plane (Fig. \ref{fig:latticeDistortionsZrP}) are displaced asymmetrically by \unit[0.21]{Å} and \unit[0.09]{Å} towards the proton, while the ions in the [010] plane (Fig. \ref{fig:latticeDistortionsZrPperp}) are displaced \unit[0.06]{Å} away. 

The oxygen ions on the cell boundary are also displaced differently. There is now an expansion of 
the supercell in the $x$-direction while in the $y$ and $z$-directions there is still a contraction but not as large as for the vacancy. This leads to the defect induced strain tensor for the proton
\eqs
\lambda_{\text{H}} = 
\begin{pmatrix}
0.009	& 0        	& 0     \\
0	& -0.066	& 0     \\
0	& 0        	& -0.010     \\
\end{pmatrix}
\label{eq:strainTensorP}
\eeqs
where the different components indicate an anisotropic strain in all three directions. Since the boundary ions are displaced outwards in the $x$-direction the corresponding strain tensor component is positive. Despite this expansion, the contractions in the other two directions lead to a negative chemical expansion coefficient of 
\eqs
\beta_{\text{H}} = -0.066
\eeqs
Similar to the case for the oxygen vacancy the lattice contracts, although the magnitude of the contraction is considerably smaller for the proton. This result agrees qualitatively with the Shannon radius of the hydroxide ion which is slightly smaller than the radius of oxygen ion \cite{ShannonActaCrystA32_1976}.

\subsection{Acceptor dopants and dopant association\label{sec:Dopants_and_dopant_association}}

We have also considered the effect of dopants and dopant association. In Tab. \ref{tab:1:dopant_association} we show the
resulting chemical expansion coefficient from a single dopant $\beta_D$ in a $2 \times 2 \times 2$ supercell for four different dopant species. This corresponds to a defect concentration of 12.5\%, which is within the range of typical concentrations (10-20\%) for real materials. The cubic structure is maintained in all cases.
The trend for the chemical expansion follows the same trend as the ionic radii by Shannon \cite{ShannonActaCrystA32_1976}, i.e. Y > In > Sc > Ga. 

It is known that there is an attractive interaction both between dopants and vacancies and between dopants and protons \cite{bjorketun_structure_2007} and dopant association is expected to be present in the real material \cite{yamazaki_proton_2013}. The expansion caused by the vacancy and the proton may therefore change if these defects are associated with a dopant atom. We have determined the chemical expansion for a vacancy and a proton with a neighbouring dopant, denoted $\beta_\text{v|D}$ and $\beta_\text{H|D}$ respectively, and the result is given in Tab. \ref{tab:1:dopant_association}. 
The resulting $\lambda$ tensor is still diagonal for the dopant-vacancy system while it becomes slightly distorted with off-diagonal components for the dopant-proton systems.

It is seen from the data in Tab. \ref{tab:1:dopant_association} that  the contraction due to the formation of the vacancy increases with increasing dopant radius. 
The same trend is observed for the proton interstitial. 
For the largest dopant Y, the distance between the B cation and the neighbouring oxygen ion increases in doping from \unit[2.12]{Å} for Zr-O to 2.22 Å for Y-O, consistent with experimental data \cite{giannici_long-range_2011}. 
This difference is less than what is obtained from the Shannon radii (\unit[0.18]{Å}) and the doped system is under compressive strain close to the dopant. 
When the vacancy or proton is added the strain can partly be reduced and the contraction becomes larger compared with the undoped system.

\begin{table}[hbt]
\begin{center}
\caption{Chemical expansion coefficients for doped and undoped \bzo\ in $2\times2\times2$ supercells. 
The results for the undoped $4\times4\times4$ supercell are included for comparison. 
The second column shows the ionic radius as tabulated by Shannon \cite{ShannonActaCrystA32_1976} and the third column the dopant chemical expansion. The ionic radii for the dopants should be compared with the radius of 0.72~Å of the Zr ion.
Column 4 shows the expansion coefficient for the trapped vacancy relative to the doped supercell and column 5 the corresponding coefficient for the trapped proton. }
\label{tab:1:dopant_association}
\begin{threeparttable}
\begin{tabular*}{\columnwidth}{@{\extracolsep{\stretch{1}}}lllll}
\hline
	&$r_\text{S}$ (Å) &$\beta_\text{D}$ 	&$\beta_\text{v|D}$ 	&$\beta_\text{H|D}$ \\	\hline
undoped\tnote{a}	& 		& 			&$-0.230$ 	&$-0.066$	\\
undoped\tnote{b}	&		& 			&$-0.246$ 	&$-0.076$	\\	
Y 			&$0.90$ 	&$\phantom{-}0.241$ 	&$-0.278$ 	&$-0.103$	\\	
In 			&$0.80$ 	&$\phantom{-}0.168$ 	&$-0.269$ 	&$-0.089$ 	\\	
Sc 			&$0.745$ 	&$\phantom{-}0.096$ 	&$-0.252$ 	&$-0.082$ 	\\	
Ga 			&$0.62$ 	&$-0.000$ 		&$-0.242$ 	&$-0.074$ 	\\	
\hline
\end{tabular*}
\begin{tablenotes}
\item[a]  $4 \times 4 \times 4$
\item[b]  $2 \times 2 \times 2$
\end{tablenotes}
\end{threeparttable}
\end{center}
\end{table}

\section{Discussion\label{sec:discussion}}

\subsection{Size of the oxygen vacancy}
We find that the lattice contracts when the vacancy is created. To quantify this effect the concept of a radius $r_\text{v}$ of a spherical vacancy has been introduced for fluorite structured oxides \cite{hong_lattice_1995,MarrocchelliAFM12,Chatzichristodoulou_size_2014} as well as for perovskites \cite{KneeGrande_JACS_97,Chatzichristodoulou_size_2014}. In the present case we find an anisotropic distorsion which is more correctly described by an ellipsoid rather than a sphere. For the change of the lattice constant we get in the $x$-direction $\Delta a_x = a_0 x_{\text{v}} \lambda_{xx}$,
where $x_{\text{v}}$ is the vacancy concentration and $\lambda_{xx}$ is the first of the three diagonal components of the tensor $\lambda$. In the $y$ and $z$-directions we get the same change $\Delta a_y = \Delta a_z$, 
with $\Delta a_y = a_0 x_{\text{v}} \lambda_{yy}$. 

How to model the distortion of the \bzo\ lattice in a consistent way using the concept of radii is not at all obvious. \bzo\ is cubic with a Goldschmidt ratio close to unity. Following Andersson \etal \cite{KneeGrande_JACS_97} we introduce an effective anion radius as 
\begin{equation}
\bar{r}_{\text{O},\alpha} = \tfrac{1}{3}\left[(3-x_{\text{v}}) r_{\text{O}} +x_{\text{v}} r_{\text{v},\alpha} \right] = r_{\text{O}} + \tfrac{1}{3}x_{\text{v}} \left(r_{\text{v},\alpha} - r_{\text{O}}\right)
\end{equation}
where $r_\text{O}$ is the radius of the oxygen ion. The subscript $\alpha$ denotes the three different directions $x$, $y$ and $z$ and indicates that the effective radius can differ in the different directions. In the present case we can write the lattice constant as $a_x = 2 (r_{\text{B}} + \bar{r}_{\text{O}})$ in the $x$-direction and $a_y = \sqrt{2} (r_{\text{A}} + \bar{r}_{\text{O}})$ for the
$y$ and $z$-directions, where $r_\text{A}$ and $r_\text{B}$ are the radii of the A-site (barium) and B-site (zirconium), respectively. The change of the lattice constants becomes  $\Delta a_x = 2x_{\text{v}} (r_{\text{v},x} - r_{\text{O}})/3$ and $\Delta a_y = \sqrt{2} x_{\text{v}} (r_{\text{v},y} - r_\text{O})/3$ and hence
\begin{align}
       r_{\text{v},x} &= r_\text{O} + \tfrac{3}{2} a_0 \lambda_{xx} \\
       r_{\text{v},y} &= r_\text{O} + \tfrac{3}{\sqrt{2}} a_0 \lambda_{yy}
\end{align}
Using the Shannon radius 
$r_{\text{O}}=\unit[1.40]{\text{Å}}$ \cite{ShannonActaCrystA32_1976} 
we get for the semi-principal axes of the ellipsoid: 
$r_{\text{v},x}=\unit[1.18]{\text{Å}}$ and $r_{\text{v},y} = r_{\text{v},z} = \unit[0.58]{\text{Å}}$. 
This can be compared with the value $r_\text{v} = \unit[1.18]{\text{Å}}$ 
used by Andersson \etal \cite{KneeGrande_JACS_97} in order to obtain an adequate fit to their experimental data.

The size of the oxygen vacancy has also been the subject of interest in studies of other perovskite oxides. Freedman \etal \cite{Freedman_PRB_2009} have performed atomistic modelling of SrTiO$_3$ using a shell-potential model by Akhtar \etal \cite{akhtar_computer_1995}. They found that the vacancy expands in the $x$-direction while contracts in the other two resulting in approximately no change in volume, i.e.~$\beta_{\text{v}}\approx0$. We have performed calculations for \bzo\ with the same type of potential using parameters taken from the work by Stokes and Islam \cite{stokes_defect_2010}. Like Freedman \etal we found that $\beta_{\text{v}}\approx0$. 

The size of the charged oxygen vacancy has been found to be smaller than the oxygen ion also in fluorite structured oxides. Hong and Virkar found the radius of the vacancy to be \unit[1.164]{Å} and \unit[0.993]{Å} in CeO$_2$ and ZrO$_2$ respectively \cite{hong_lattice_1995}. Marrocchelli \etal found the radius to be \unit[1.169]{Å} and \unit[0.988]{Å} for the same materials using both theoretical modelling and experimental techniques \cite{MarrocchelliAFM12,marrocchelli_charge_2012}.

\subsection{Chemical expansion due to hydration}

With chemical expansion coefficients for the oxygen vacancy and the proton interstitial we can now consider the chemical expansion due to hydration. As can be seen from Eq. \eqref{eq:hydration}, hydration of the material corresponds to replacing one oxygen vacancy with two proton interstitials. The chemical expansion per incorporated proton is thus given by
\eq
\beta_{\text{hydr}} = \beta_{\text{H}}-\tfrac{1}{2}\beta_{\text{v}}
\label{eq:volume_expansion}
\eeq

In Tab. \ref{tab:2:dopant_association_hydration} we show the result for both supercell sizes. The chemical expansion is positive, which corresponds to an expansion of the lattice. 
The chemical expansion for the two different supercell sizes differs by about 4\%, indicating an interaction effect due to the finite defect concentration.

The formula in Eq. \eqref{eq:volume_expansion} neglects the explicit effect from the dopants. If association between the dopant and
the vacancy and/or proton is present the chemical expansion will change. However, we found in section \ref{sec:Dopants_and_dopant_association} that the effect is quite small. 
If we assume that the vacancy is associated with a single dopant, as a nearest neighbouring defect, the expression in Eq. \eqref{eq:volume_expansion} should be modified to
\eq
\beta^{(1)}_{\text{hydr}} = \beta_{\text{H}} \;\; -\tfrac{1}{2}\beta_{\text{v|D}} 
\label{eq:free_protons}
\eeq
If also the proton is trapped at a dopant site Eq. \eqref{eq:volume_expansion} should be modified to
\eq
\beta^{(2)}_{\text{hydr}} = \beta_{\text{H|D}}-\tfrac{1}{2}\beta_{\text{v|D}}
\label{eq:trapped_protons}
\eeq
In Tab. \ref{tab:2:dopant_association_hydration} the results are summarised. The effect from the dopants is not pronounced. When only vacancy-dopant association is included the chemical expansion increases with increasing dopant radius.
If also proton-dopant association is included the expansion is rather similar to the undoped system, except for the yttrium doped system where the expansion is reduced.  

\begin{table}[!b]
\begin{center}
\caption{Chemical expansion due to hydration for undoped and doped \bzo\ in $2\times2\times2$ supercells. 
The results for the undoped $4\times4\times4$ supercell are included for comparison. 
The second column shows the expansion according to Eq. \eqref{eq:volume_expansion}. The third column shows the expansion for trapped vacancies (Eq. \eqref{eq:free_protons}) and the fourth column for trapped vacancies and protons (Eq. \eqref{eq:trapped_protons}).}
\label{tab:2:dopant_association_hydration}
\begin{threeparttable}
\begin{tabular*}{\columnwidth}{@{\extracolsep{\stretch{1}}}lllll}
\hline
	&$\beta_\text{hydr}$		&$\beta_\text{hydr}^\text{(1)}$	&$\beta_\text{hydr}^\text{(2)}$	\\ \hline
undoped\tnote{a}	&$0.049$	&		&		\\ 
undoped\tnote{b}	&$0.047$	&		&		\\
Y			&		&$0.063$	&$0.037$	\\
In			&		&$0.059$	&$0.046$	\\
Sc			&		&$0.050$	&$0.044$	\\
Ga			&		&$0.045$	&$0.047$	\\
\hline						
\end{tabular*}
\begin{tablenotes}
\item[a]  $4 \times 4 \times 4$
\item[b]  $2 \times 2 \times 2$
\end{tablenotes}
\end{threeparttable}
\end{center}
\end{table}

As mentioned previously in the introduction, lattice expansion due to hydration in \bzo\ has been seen experimentally for several different acceptor dopants and dopant concentration levels. Based on the measured results we can extract the relative lattice expansion $(\Delta a/a_0)$ which can be used to calculate the chemical expansion coefficient according to
\begin{equation}
\beta_{\text{hydr}} = 3\frac{\Delta a/a_0}{x_{\text{H}}}
\end{equation}
where $x_{\text{H}}$ is the proton concentration.

Kreuer \cite{Kreuer_AnnRevMatRes2003} measured the relative lattice expansion in 15\% Y-doped \bzo\ and found it to be 0.0040. An almost fully hydrated sample with a proton concentration of 0.14 (data extracted from Fig.~2a) has been reported previously by Kreuer \etal \cite{kreuer_proton_2001}. Kinyanjui \etal \cite{EricssonSSI225_2012} studied the hydration of 50\% In-substituted \bzo. The hydration gave rise to an expansion of 0.0107 (data extracted from Fig.~2), although the proton concentration was only 0.40 (data extracted from Fig.~3). Furthermore, Hiraiwa \etal \cite{HiraiwaJAmCeramSoc96_2013} performed a study on 20\% Y-doped \bzo. They found a relative expansion of 0.0019 (data extracted from Fig.~3) which is a bit lower than the previously discussed values. They do not present any results related to the degree of hydration although it is quite reasonable to assume that the material was not fully hydrated based on the small expansion. This work was followed by a study of 20\% doped \bzo\ where four 
additional dopants where considered, namely Sc, Sm, Eu and Dy \cite{han_dopant_2014}. Again, the relative lattice expansions were quite small, ranging from 0.0007 to 0.0016 (data extracted from Tab. 2). Besides the possibility of an incomplete hydration the small expansion values could also be related to the fact the these dopants have a tendency to occupy the A-site instead of the B-site  \cite{han_site_2012,han_dopant_2014}, which reduces the concentration of oxygen vacancies and consequently the maximum proton concentration. Additionally, the authors found results indicating that the Dy ions could change oxidation state. This is associated with a change in the ionic radius, which would affect the chemical expansion. Finally, in a recent study by Andersson \etal \cite{KneeGrande_JACS_97}  chemical expansion was observed for several different Y-doped samples of \bzo\ with dopant concentrations of 5\%, 10\% and 20\%. The relative expansion was 0.0015 for both the 5\% and 10\% doped samples while it was 0.
0034 for the sample with 20\% yttrium. TGA data presented in the paper indicate that the 5\% doped sample was fully hydrated while the samples with 10\% and 20\% dopants were not.

The chemical expansion coefficient ($\beta_{\text{hydr}}$) has been calculated for the discussed experimental results and are summarised in Tab. \ref{tab:chemicalexpansion} together with the relative lattice expansion and the degree of hydration. All these values for $\beta_{\text{hydr}}$ indicate an expansion that is consistent with our result in Tab.  \ref{tab:2:dopant_association_hydration}. 

It seems that we may underestimate the expansion based on the comparison with the data by Kreuer \etal \cite{Kreuer_AnnRevMatRes2003,kreuer_proton_2001} and Kinyanjui \etal \cite{EricssonSSI225_2012}. 
One uncertainty is the approximation for the exchange-correlation functional. 
The data in this work were obtained using the GGA/PBE approximation. However, some test calculations using the standard local density approximation (LDA) indicate that the results are rather insensitive to the particular choice of exchange-correlation approximation. 
Finite temperatures and thermal fluctuations may also influence the chemical expansion. These effects are not included in this work since our data are determined from fully relaxed configurations, which correspond to zero Kelvin. 
For the proton quantum fluctuations can also play a role.

\begin{table}[bt]
\begin{center}
\caption{Experimental values of the relative lattice expansion coefficient due to hydration ($\Delta a/a_0$) and the corresponding chemical expansion coefficient ($\beta_\text{hydr}$) for acceptor-doped \bzo\ with various dopant species and concentrations. $x_{\text{A}}$ denotes the nominal dopant concentration and $x_{\text{H}}$ the estimated actual proton concentration.}
\label{tab:chemicalexpansion}
\begin{tabular*}{\columnwidth}{@{\extracolsep{\stretch{1}}}llllll}
\hline
Dopant    	& $x_{\text{A}}$	& $\Delta a/a_0$ & $x_{\text{H}}$ &  $\beta_\text{hydr}$	&    Ref.            \\ \hline
Y		& 0.15 & 0.0040 & 0.14 & 0.086				& \cite{Kreuer_AnnRevMatRes2003,kreuer_proton_2001} \\
In		& 0.50 & 0.0107 & 0.40 & 0.080				& \cite{EricssonSSI225_2012}     \\
Y		& 0.20 & 0.0019 & $<0.20$	& $>0.028$		& \cite{HiraiwaJAmCeramSoc96_2013}\\
Sc 	   	& 0.20 & 0.0014 & $<0.20$	& $>0.021$		& \cite{han_dopant_2014}        \\
Sm		& 0.20 & 0.0010 & $<0.20$	& $>0.016$		& \cite{han_dopant_2014}		   \\
Eu	    	& 0.20 & 0.0016 & $<0.20$  	& $>0.024$		& \cite{han_dopant_2014}        \\
Dy	    	& 0.20 & 0.0007 & $<0.20$  	& $>0.010$		& \cite{han_dopant_2014}        \\
Y		& 0.05 & 0.0015 & $\approx0.05$ & $\approx0.090$	& \cite{KneeGrande_JACS_97}    \\
Y 	   	& 0.10 & 0.0015 & $<0.10$ 	& $>0.045$		& \cite{KneeGrande_JACS_97}    \\
Y 	   	& 0.20 & 0.0034 & $<0.20$	& $>0.051$		& \cite{KneeGrande_JACS_97}    \\
\hline
\end{tabular*}
\end{center}
\end{table}

\section{Conclusions\label{sec:conclusions}}

Density functional theory (DFT) is used to compute the defect induced strain tensor $\lambda$ for two point defects in barium zirconate: a +2 charged oxygen vacancy and a proton interstitial (forming a hydroxide ion).
The defects have been considered both independently and in association with dopant ions.
The tensor $\lambda$ provides a natural generalisation of the chemical expansion coefficient to anisotropic deformations. 
 
We find that the lattice contracts both when the vacancy and the proton are inserted, indicating that both the vacancy and the hydroxide ion are smaller compared with an oxygen ion.
We also find that the effect from dopant association is quite small. 
The contraction for the vacancy is considerably larger than for the proton and the net effect in hydration, when a vacancy is filled and two protons are added, is an expansion, consistent with the experimental findings.

\section*{Acknowledgements}

We thank Christopher Knee at the Department of Chemical and Biological Engineering, Chalmers University of Technology for instructive comments and acknowledge the Swedish Energy Agency for financial support (Project number: 36645-1). Computational resources have been provided by the Swedish National Infrastructure for Computing (SNIC) at Chalmers Centre for Computational Science and Engineering (C3SE) and National Supercomputer Centre (NSC).

\bibliographystyle{model1a-num-names}
\bibliography{ChemicalExpansion}

\end{document}